\documentclass{acm}

\usepackage{times}  
\usepackage{array, graphics, graphicx, color, url, xspace, multirow, rotating, epsfig, amsmath, amsfonts}
\usepackage{breakurl}
\usepackage{wrapfig}
\usepackage{tikz}
\usepackage{epstopdf,subfig}
\usepackage[format=plain,labelfont=bf,font=small]{caption}
\usepackage{ragged2e,flushend,enumitem,listings}
\usepackage{blindtext}
\usepackage{acronym} 
\usepackage{hyperref, cleveref}
\usepackage{multirow}
\usepackage{svg}
\usepackage{multicol}

\usepackage{appendix}
\usepackage{tabularx}
\usepackage[small,compact]{titlesec}
\setitemize{noitemsep,topsep=0pt,parsep=0pt,partopsep=0pt}

\usepackage{pifont}%

\usepackage{listings}[mathescape]
\usepackage{color}
\usepackage{outlines}

\hypersetup{pdfstartview=FitH,pdfpagelayout=SinglePage}

\newcommand\paragraphb[1]{\noindent{\bf{#1}}}
\newcommand\paragraphi[1]{\noindent\emph{#1}}
\newcommand\pghi[1]{\paragraphi{#1}}
\newcommand\pb[1]{\paragraphb{#1}}
\renewcommand\pi[1]{\paragraphi{#1}}

\newcommand{\bi}{\begin{itemize}}
\newcommand{\ei}{\end{itemize}}

\newcommand{\eg}{\emph{e.g.,}\xspace}

\newcommand{\eat}[1]{}

\newcommand{\squishlist}{
  \begin{list}{$\bullet$}{
    \setlength{\itemsep}{0pt}       \setlength{\parsep}{3pt}
    \setlength{\topsep}{3pt}        \setlength{\partopsep}{0pt}
    \setlength{\leftmargin}{1em}    \setlength{\labelwidth}{1em}
    \setlength{\labelsep}{0.5em} } }

\newcommand{\squishend}{
  \end{list} }
\newcommand{\squishlistend}{
  \end{list} }

\definecolor{codegray}{rgb}{0.5,0.5,0.5}
\lstdefinestyle{bashStyle}{
    commentstyle=\color{blue},
    numberstyle=\tiny\color{codegray},
    basicstyle=\ttfamily\scriptsize,
    breakatwhitespace=false,         
    breaklines=true,                 
    captionpos=b,                    
    keepspaces=true,                 
    numbers=left,                    
    numbersep=5pt,                  
    showspaces=false,                
    showstringspaces=false,
    showtabs=false,                  
    tabsize=2,
    keywordstyle=\bfseries,
}

\graphicspath{{./dia/}}
\begin{document}

\title{From Internet of Things to Internet of Data Apps}
\author{Silvery Fu and Sylvia Ratnasamy \\ UC Berkeley}
\maketitle

\begin{abstract}
We introduce the Internet of Data Apps (IoDA), representing the next natural progression of the Internet, Big Data, AI, and the Internet of Things. Despite advancements in these fields, the full potential of \emph{universal data access} - the capability to seamlessly consume and contribute data via data applications - remains stifled by organizational and technological silos. To address these constraints, we propose the designs of an IoDA layer borrowing inspirations from the standard Internet protocols. This layer facilitates the interconnection of data applications across different devices and domains. This short paper serves as an invitation to dialogue over this proposal.
\end{abstract}

\section{Introduction}
\label{sec:intro}

Three decades ago, Mark Weiser predicted a future where technology would become a seamless part of our lives~\cite{weiser1994ubiquitous}. This prediction is now reflected in the numerous data sources that are part of our everyday existence, such as mobile devices, the Internet of Things (IoT), wearable tech, connected vehicles, and smart infrastructures~\cite{home-statista,fitbit,lu2014connected,smartthings,smart-building}. These sources aren't restricted to tangible hardware or software applications but can also include conceptual collections, such as the data produced by entire buildings or campuses. Concurrently, we've seen a rise in data-driven applications designed to interact with these data sources, offering solutions for navigation~\cite{google-maps}, food delivery~\cite{doordash}, ride-hailing~\cite{uber}, event planning~\cite{envoy}, and home automation~\cite{alexa}.

Despite the ubiquity of both data sources and applications, \emph{universal data access} – the ability for everyone to consume and contribute data – is not yet a reality. To understand this, consider how devices connect to networks. When you visit a new city, your smartphone automatically connects to the local network. At a restaurant, you join the public Wi-Fi without any issues. This ease of connectivity is not mirrored in data access. Today, consuming diverse data types – real-time traffic updates, air quality readings, local event schedules, public safety notices, or e-bike or scooter availability – often means toggling between separate, disconnected data apps. This fragments user experience, and leads to loss of valuable insights and time, making it difficult to leverage the full potential of the available data.

Similarly, these data apps/platforms are typically limited in their ability to incorporate user data contributions, preventing potentially valuable inputs from being used effectively. Consider, for example, if your smartphone could share real-time network quality data across different city areas. This data could then be easily collated, aiding more informed decisions about connectivity. Likewise, if connected vehicles could share real-time traffic and road condition data to enhance traffic management. Or, imagine if local businesses could contribute data about current wait times at restaurants or the availability of rental bikes at specific locations. 

Existing data apps like Google Maps~\cite{google-maps}, while effective at aggregating certain types of data, are not designed to provide a universal data access experience. Because these data aggregators face the same data access limitations, hindering their ability to evolve and provide universal access. For example, they cannot dynamically integrate information from varied sources according to users' changing needs, As interfaces to access data, these data apps' limitations ultimately impact the user's ability to gain universal data access.

We observe that achieving universal data access is complicated by two primary obstacles. The first one is the \emph{semantic gap}, referring to the difference between the raw data generated by sources and the processed, ready-to-use data required by applications. Transforming raw data into a format usable by applications – a process we refer to as \emph{data curation} – involves various steps such as filtering, transformation, and aggregation. The second obstacle is the \emph{technical stack gap} or (stack gap for short), which prevents seamless data access between data apps due to the lack of interoperability across different technological stacks supporting/behind the data apps. This can include difficulties in discovering, authenticating, and authorizing data access and actual exchanges of data across the stacks. In fact, modern data infrastructures (\S\ref{sec:vision}), while designed to address the semantic gap, often exacerbate the stack gap by increasing complexity and hindering interoperability. As a result, even when other potential obstacles such as policies and privacy regulations permit data access, the fundamental challenges posed by these two gaps can still prevent data applications from easily accessing each other's data.

In this paper, we argue that for universal data access to become a reality, two things need to happen. First, we must be able to carry out \emph{data curation on an Internet-scale} to bridge the semantic gap. Second, this curated data should be exchangeable between data applications with the stack gap out of the way. How to achieve the two goals? For the semantic gap, our conjecture (\S\ref{sec:vision}) is that addressing the semantic gaps has now become feasible and scalable due to the advancements in the field of machine learning, particularly the introduction of solutions like GPT/LLM~\cite{llm}. These advanced AI models are capable of understanding, interpreting, and manipulating data in sophisticated ways that were previously unattainable~\cite{meta-lm-di}. For the stack gap, we argue we should draw inspirations from the design principles of the Internet~\cite{clark}, particularly in how it achieves interconnection across diverse networks. In much the same way, we need to focus on creating an infrastructure that can connect various data apps and their ``app domains'', crossing their stack gaps.

With these insights, we propose the Internet of Data Apps (IoDA)\footnote{Pronounced as U-DA. As in, \emph{IoDA provides universal data access.}}---a universal data access layer aimed at furthering Weiser's vision by enabling ubiquitous data access at Internet-scale. In this paper, we outline the design principles for IoDA and present a technical design that aligns with these principles. Our vision is to create a world where individuals and systems can seamlessly consume and contribute data, representing a natural progression of the Internet and the Internet of Things as we know and rely on them today.
\section{The Vision}
\label{sec:vision}

In this section, we present the trends/vision, examples, and requirements of IoDA.

\subsection{Primer and Trends}
\label{subsec:trend}

We observe the following emerging trends, each contributing to the contexts based on which our Internet of Data Apps vision will develop:
\squishlist
\item \pb{Proliferation of mobile and IoT devices:} The number of mobile and IoT devices~\cite{alexa,apple-watch,fitbit,apple-home} is continuously increasing, with users relying on these devices for diverse daily tasks. This trend is facilitating the ubiquity of data access, with both users and their devices acting as data sources. However, a gap exists between the data generated and its accessibility and usability. Today, while we can view varied data types on our phones and our devices can produce a wealth of data, the ability to aggregate, process, and use this information effectively is limited.
\item \pb{Big data processing frameworks:} The evolution of big data processing frameworks, such as Databricks~\cite{databricks} and Snowflake~\cite{snowflake}, alongside the emergence of data integration tools like Fivetran~\cite{fivetran} and Airbyte~\cite{airbyte}, has transformed how we handle and manipulate large data sets. These tools allow efficient data management, which are complementary and instrumental in supporting universal data access.
\item \pb{Maturity of AI and language models:} Advances in AI, particularly language models, have revolutionized data automation and processing. Models like GPT have shown remarkable competence in natural language tasks~\cite{llm}, while others have exhibited the capability to process complex database queries~\cite{li2023can} and carry out data processing tasks~\cite{meta-lm-di}. These advancements provide a robust foundation for data curation, an integral part of achieving IoDA's vision.
\item \pb{Non-universal data access:} Despite the aforementioned advancements, universal data access remains elusive. In particular, today's data landscape is fragmented, with data sources and applications often functioning in isolation. This lack of interoperability restricts data's potential, hindering its accessibility and usability. 
\squishend

\pb{\underline{IoDA vision:}} With these trends as context, we argue that a universal data access layer should emerge. It should leverage the ubiquity of data sources and the advancements in AI and data processing frameworks to address the challenge of non-universal data access. The goal is to create an infrastructure that facilitates bridging semantic gap through AI-enabled data curation while eliminating the stack gap by promoting interoperability across diverse data applications. In doing so, IoDA seeks to actualize the promise of universal data access, enabling individuals and systems to seamlessly consume and contribute data at an Internet-scale.

\subsection{The Real-world Drives for IoDA}
\label{subsec:benefit}

In what follows, we explain the more tangible benefits of the IoDA from the perspectives of end-users and enterprises.

\pb{\#1: Benefits for enterprises.} The promise of universal data access bears significant implications for enterprises, especially in optimizing operational efficiency, resource utilization, and cost-saving. Consider the context of smart buildings, a market projected to surpass \$121.6 billion by 2026~\cite{markets,venturebeat}. A key application within this ecosystem is space analytics~\cite{venturebeat,density,comfy}, which enables managers to make data-driven decisions by collecting and analyzing historical and real-time data on space utilization and occupancy patterns. Currently, space analytics requires either a dense network of expensive and maintenance-intensive sensors or a reliance on booking systems that infer utilization via user interactions. With universal data access, a paradigm shift towards a ``self-serve'' space becomes possible, reducing costs and improving resource utilization. In this scenario, building managers can leverage data from tenants' BYOD devices (\eg laptops, smartphones, wearables), with user consent, to enhance the accuracy of space analytics. Tenants' devices contribute real-time occupancy data, reducing the need for dedicated sensor installations and offering more precise occupancy insights. 

Taking this example further, smart buildings could also connect with the data ecosystem of a city, receiving real-time data about factors such as traffic conditions, weather forecasts, and event schedules~\cite{envoy}. These data can be utilized to further optimize building operations like energy management and to improve tenant services. More broadly, for enterprises and business, the potential for having universal data access goes beyond cost and efficiency. By facilitating real-time data exchange with different business entities, \eg from transport services to local restaurants – businesses can unlock new revenue streams. For example, by sharing real-time data about their customers' preferences and behaviors (with proper consent), businesses can offer personalized marketing and services, creating additional revenue opportunities.

\pb{\#2: Benefits for end-users.} Imagine a day in the life of a city resident in a world with universal data access. As she \textbf{(1)} wakes up in the morning, her smart home has already prepared the perfect environment by adjusting the lighting, temperature, and even playing her favorite music, having seamlessly accessed data from her sleep tracking app. \textbf{(2)} Upon heading to the gym, her fitness tracker automatically shares her health stats and exercise preferences with the gym's network. This data enables personalized workout recommendations and real-time progress tracking, enhancing her fitness routine. Meanwhile, the tracker collects data about the gym's equipment usage, contributing to a community dataset that the gym uses to optimize its facilities. \textbf{(3)} She then visits a shopping mall, where her smartphone connects to the mall's network. The mall system, equipped with data about her past purchases and preferences, guides her to stores with items she might like, improving her shopping experience. As she moves around, her phone shares anonymized data about her shopping behaviors, aiding the mall in improving store placements and customer service. \textbf{(4)} Taking public transit to work, her phone communicates with the transit system, providing real-time updates on schedules and seat availability. In return, her device shares usage pattern data, assisting the transit authority in improving routes and schedules. \textbf{(5)} Finally, at her office, walking into a meeting room automatically connects her smartphone to the room's systems, allowing her to control the environment and be informed of the room's schedule. Her device, in return, provides data about room usage and energy consumption patterns to the building management system.

\pb{\underline{Summary.}} To summarize, we argue that IoDA can offer enterprises significant improvements in operational efficiency, cost savings, and potential revenue growth by establishing an interconnected data ecosystem. Meanwhile, from a user perspective, IoDA can create a highly personalized, efficient, and responsive living and working environment.

\subsection{Technical Requirements for IoDA}
\label{subsec:requirement}

We argue that to achieve universal data access, the following requirements are \emph{necessary}:

\squishlist
\item \textbf{\#1 Universal accessibility:} Data should be accessible from any location, domain, and on any device, when allowed so. Similar to how the Internet provides universal access to information, universal data access implies that all relevant data is readily available when/wherever it is needed.
\item \textbf{\#2 Interoperability:} Data from different sources should be readily compatible or easily made so for access. This means that data apps from disparate organizations/domains should be able to exchange and use the information.
\item \textbf{\#3 Real-Time and opportunistic access:} Depending on the nature of the data and its use, access should be allowed to happen in real-time or near real-time. There should not be significant latency that would render the data outdated and therefore less useful. Besides, data apps should be allowed for opportunistically access data, adapting to unexpected data sources and data consumers/other data apps.
\item \textbf{\#4 Bi-directionality:} Universal data access is not only about consuming data but also about contributing data. Data apps should be able to easily contribute data to the pool of data, as well as consume data from it.
\item \textbf{\#5 Ease of use:} Accessing data should be as simple and intuitive as connecting to and using a network. It should not require specialized knowledge or complex procedures.
\item \textbf{\#6 Security and privacy:} Given the sensitive nature of many types of data (\eg PII data) that might be accessed and shared, it's imperative that senew curity measures are in place to protect data integrity and confidentiality.
\squishend

\noindent We'll revisit these requirements as we describe the designs of our IoDA proposal (\S\ref{sec:proposal}).
\section{A Design Proposal}
\label{sec:proposal}

This section present the design requirements and principles (\S\ref{subsec:principle}) and the approach meeting these principles (\S\ref{subsec:approach}).

\subsection{Principles}
\label{subsec:principle}

\begin{figure}[t]
     \centering
     \footnotesize
     \includegraphics[width = 0.47\textwidth]{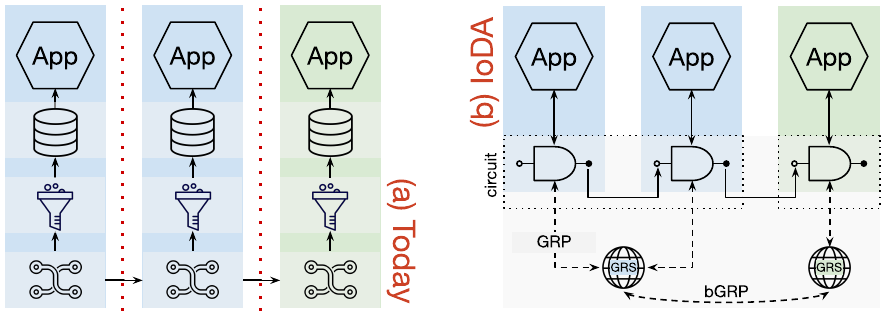}
     \caption{IoDA architecture vs. today.}
     \label{fig:arch}
\end{figure}

The highest guiding principles of IoDA is striving for simplicity - both in the sense of arriving at a simple design that is easy-to-adopt as well as the simplifying the applications and users it aim for. Data access should be easy and intuitive to be ubiquitous, and the processes of contributing and consuming data should be as straightforward as uploading and downloading packets from the Internet, from the perspective of data apps. We propose the following design principles:

\squishlist
\item \pb{Design Principle \#1: Unified abstractions for data access:} A cornerstone of IoDA's design is the concept of unified data app abstractions, whereby a data app is both a data consumer and a data contributor. This role combination eliminates technological barriers that could potentially trap data in silos. From a technical standpoint, the data app abstraction in IoDA serves as both a data source and a data app, providing a unified interface for data sources (like an IoT sensor), data consumers (such as a dashboard), and entities that perform both functions (\eg a smart light adjusting to power settings while roporting energy readings). This unification simplifies system operations by allowing the use of a consistent set of mechanisms across all layers.
\item \pb{Design Principle \#2: Treating metadata as data but decoupling it from actual data:} This principle addresses the issue of metadata, which is integral for the organization, access, and interpretation of data. In IoDA, metadata is treated as data, enabling its independent management and processing. However, it's important to decouple metadata from the actual data to ensure their independent evolution and operation and to prevent unnecessary coupling.
\item \pb{Design Principle \#3: Allowing data app composition at any layer and across any domains:} IoDA should support data app composition at any layer, and the system mechanisms should not hinder this functionality. This design allows multiple data apps or domains to be federated or combined, irrespective of their layers or app domains. 
\item \pb{Design Principle \#4: Decoupling specifications from executions on all layers:} Specifications and executions should be independent at all system levels in IoDA. This principle has two main advantages. First, it enables individual app providers/domains to choose their implementations of the data processing systems without affecting the shared \emph{semantics} of data processing and sharing. For example, while one domain might use Databricks for data processing, another might employ Snowflake; however, the overall semantic interpretation of the data processing pipelines remains consistent. Second, such decoupling also enables optimizations and verifications at all levels, enhancing the system's overall efficiency and reliability.
\squishend

\subsection{Approach}
\label{subsec:approach}

To help elaborate on our design, we define the following concepts within IoDA: (i) \emph{data app:} A data app consumes and provides data, acting as the instantiation of data access. A data app could be an end-user interface, a UI, or a dashboard; (ii) \emph{data domain:} Each domain is owned by different domain owners, with clear boundaries for data access and policy enforcement; and (iii) \emph{metadata:} This encompasses data schemas, access endpoints, and data descriptions, which provide information for locating and accessing data. (iv) \emph{data:} The actual data records. Our design below focuses on IoT-like data, which can be structured or semi-structured.

\pb{Overview.} The design of the IoDA layer comprises the following components: (a) \emph{Data gateway (gate):} A data gateway curates data, making it ready for consumption by the data app and other data gates. It consumes data from defined data sources and exports processed data. Serving as both a specification interface for the data app developers and a runtime execution abstraction. (b) \emph{Wire:} Data wires are the ``conduits'' or ``pipes'' that connect data gateways and execute data movements and interconnections across multiple data gates, either within or across circuits. (c) \emph{Circuit:} Circuits are formed by interconnected gates, offering global abstraction, visibility, and optimization across these gates. Each domain can have one or multiple data circuits. (d) \emph{Gate Resolution Protocol (GRP) and Border Gate Resolution Protocol (bGRP):} These protocols facilitate data exchanges within and across domains, effectively supporting the flow of data in the IoDA.

\pb{\underline{D1: Designing for universal data abstraction.}} The main objective is to recognize the shared representation and instantiation of data access and curation. We propose the concept of a data gateway, analogous to packet routers/gateways and middleboxes that process incoming packets and forward them to other gateways.

\pb{Data gateway (gate).} IoDA represents each context with a gate. A gate consists of three components: a data store, one or more input ports (iports), and one or more output ports (oports). These three abstractions allow a gate to ingest, process, store, and export data records to/from other gates.

\squishlist
\item \pghi{Input port (iport).} An iport of a gate retrieves data from one or multiple data sources. A gate can have multiple iports, each identified uniquely. Each iport processes the ingested data with its dataflow, which contains operators (\eg sort, join) and functions (\eg sum, avg) that process a sequence of input data records and generate a sequence of output records. The derived data is written to the store.
\item \pghi{Output port (oport).} An oport exposes data records in the data store to data consumers such as other gates, apps, and users. At runtime, the oport reads data from the data store, processes it, and caches it persistently. We refer to this resulting data as the "oport view". The oport supports interfaces to query or watch the data records. A gate can have multiple oports, and entities can retrieve data by querying one or more of the gate's oports. The data schemas and access control policies are specified in each oport.
\item \pghi{Data store.} The data store is an interface to the domain's data storage choices, such as a database or an object store.
\item \pghi{Data and metadata flow.} The data flow refers to the actual data records exchanged between data apps. It can be structured, semi-structured, or free-form data. For the smart city use case, IoT data records are envisioned to be represented in JSON or JSON-like formats, which provide richer types. Besides, each gate exposes metadata about the data, such as data schemas, contextual information, and data descriptions, to facilitate data discovery.
\squishend

\squishlist
\item \pb{\underline{D2: Designing for data discovery}} The high-level goal of designing data discovery in IoDA is to discover the oports across different gates and domains. 
\item \pghi{Gate addressing.} Each gate is identified using a specific addressing format: \texttt{domain/gate/oport}. This addressing scheme allows for referencing individual output ports. The domain name can be derived from existing naming infrastructure, such as Internet domain names, public ledger addresses~\cite{tet}, or GitHub accounts or organizations~\cite{github}. It's important to note that the gate address is distinct from the network address.
\item \pghi{Gate Resolution Service.} The Gate Resolution Service (GRS) resolves gate addresses, similar to how DNS resolves domain names to internet addresses. However, the GRS differs from DNS in that it resolves the iport's data source to an oport address based on both the metadata of the source gate and the destination gate. This resolution process takes into account the specific metadata of the gates involved in order to determine the appropriate gate address for data discovery.
\squishend

\begin{figure}[t]
     \centering
     \footnotesize
     \includegraphics[width = 0.47\textwidth]{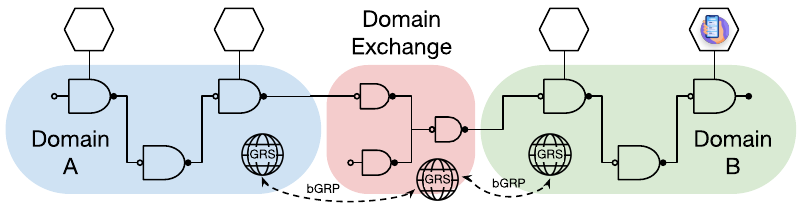}
     \caption{Data exchanges intra- and inter-domain.}
     \label{fig:arch}
\end{figure}

\pb{\underline{D3: Designing for cross-Domain access}} Cross-domain access in IoDA enables data sharing and collaboration between different domains. To support secure and controlled data exchange across domains, we propose two components:

\squishlist
\item \pb{Wire} Besides data movement, the wire component in IoDA is also responsible for gate authentication and authorization. It establishes secure communication channels between gates to ensure that data is transmitted only between authenticated and authorized gates. The wire component also enforces access control, verifying the identity and permissions of gates involved in the data exchange. 
\item \pb{Circuit} The circuit component in IoDA focuses on topology verification and enforcement. A circuit is formed by interconnected gates, representing the logical path through which data flows between domains. The circuit provides a global abstraction and visibility across gates, allowing for optimized data movement and processing. In addition, the circuit verifies the connectivity and integrity of the gates within the topology, ensuring that data can be exchanged seamlessly between domains while maintaining the required security and privacy measures.
\squishend

\pb{\underline{D4: Designing for privacy and security}} We focus on the following high-level points privacy and security:
\squishlist
\item \pghi{Access control.} IoDA incorporates access control mechanisms to regulate data access. Role-based Access Control (RBAC) is employed to determine which roles can access specific data. This ensures that only authorized entities can interact with the data.
\item \pghi{Ownership.} IoDA can improve data ownership and control. Gate operators have the authority to determine where and how context data is stored. This allows them to maintain control over their data and choose the storage systems that align with their requirements.
\item \pghi{Provenance and Governance.} IoDA enables the tracking of data provenance, capturing the sources and modifications of data. This provides transparency and accountability in data handling. Further, IoDA supports governance policies that ensure data quality and compliance with regulations. These policies enable users to define rules and restrictions on data usage to maintain privacy and meet regulatory standards.
\squishend

\pb{IoDA Deployment.} IoDA allows different deployment strategies. A common scenario involves users utilizing a lightweight client-side app to interact with gates deployed in either a cloud or on-premises environment. IoDA providers, such as SaaS companies, operate IoDA clusters comprising the runtime components and users' gates. A user's gate is hosted by a ``home provider,'' while the option to register gates with non-home providers enables gate integration across clusters. For example, a device vendor may act as a provider, running a user's phone gate as a service accessible through a phone app. When a user joins a context like a building, their device gate automatically connects with the building's gate hosted by a separate provider. Other scenarios include single-provider gate hosting or apps incorporating known/pre-configured gates, ensuring flexibility and seamless integration within an IoDA cluster.

\section{Call for Research Dialogue}
\label{sec:disc}

\pb{Why is systems and networking community well-suited for designing IoDA?} Our networking community possesses the essential characteristics necessary for designing IoDA. First, our community encompasses multiple disciplines, making it inherently multi-disciplinary~\cite{fff} (see below). Second, as experts in data communication, we have the technical expertise and insights required to address the challenges of data access and sharing. We have a deep understanding of networking techniques and perspectives, which can be leveraged to design IoDA effectively. Third, our community has a successful track record in solving interoperation problems, as demonstrated by our accomplishments in building and evolving the Internet. By applying the principles and lessons learned from previous networking advancements~\cite{clark,shenker,cerf,tbl}, we are at the vantage point to develop IoDA.

\pb{Why is collaboration with other communities crucial for IoDA's success?} The realization of IoDA relies on the collaboration with other communities too. First, the system community, encompassing big data systems, streaming systems, IoT systems, cloud computing, and edge systems, plays a pivotal role in providing the underlying infrastructure and technologies required by IoDA. Second, the database community brings expertise in data models~\cite{zed}, data integration~\cite{meta-lm-di}, query optimization~\cite{query-opt}, and data engine design~\cite{telegraph}. Their knowledge and advancements are essential for developing robust and efficient data processing and management mechanisms within IoDA. Finally, the HCI community specializing in user interface design plays a critical role in shaping the user experience of universal data apps, ensuring they are intuitive, accessible, and user-friendly.

Further, there is a pressing need to address the growing dominance of ``incumbents'' such as Snowflake~\cite{snowflake} in the data industry. To democratize data and empower users to control their own data and access, it is crucial to develop an IoDA layer that fosters data ownership and accessibility. This paper serves as an early call for research dialogue, inviting collaboration and discussions among researchers and practitioners from various communities. We are actively prototyping systems to support the IoDA layer and will share our findings and experiences in our follow-up papers.

\bibliographystyle{abbrv} 
\bibliography{ioda}

\end{document}